\documentclass[aip,jmp,amsmath,amssymb,amsfonts,reprint]{revtex4-1}

\usepackage{graphicx}

\begin{document}

\title{Control of optical properties of hybrid materials with chirped femtosecond laser pulses under strong coupling conditions}

\author{Maxim Sukharev}
\affiliation{Science and Mathematics Faculty, School of Letters and Sciences, Arizona State University, Mesa, Arizona 85212, USA}
\email{maxim.sukharev@asu.edu}
\date{\today}

\begin{abstract}

The interaction of chirped femtosecond laser pulses with hybrid materials - materials comprised of plasmon sustaining structures and resonant molecules - is scrutinized using a self-consistent model of coupled Maxwell-Bloch equations. The optical properties of such systems are examined with the example of periodic sinusoidal gratings. It is shown that under strong coupling conditions one can control light transmission using chirped pulses in a spatiotemporal manner. The temporal origin of control relies on chirps non-symmetric in time while the space control is achieved via spatial localization of electromagnetic energy due to plasmon resonances.

\end{abstract}


\maketitle

\section{Introduction}
\label{sec:introduction}
The push towards nanoscale optical elements that can operate on a femtosecond timescale is ultimately converging to plasmonic materials with their unique ability to scatter electromagnetic (EM) radiation far below the diffraction limit resulting in high local EM fields.\cite{ADMA:ADMA200700678,Gramotnev:2010aa,Schuller:2010aa} Utilization of optical properties of plasmons has opened up a wide variety of applications.\cite{Stockman:11} Specific to this paper is a merger of research in plasmonics with physics of ensembles of quantum emitters.\cite{torma2014strong} The physics of hybrid materials - systems comprised of quantum emitters (quantum dots,\cite{PhysRevLett.99.136802} J-aggregates\cite{PhysRevLett.93.036404}) optically coupled to plasmonic structures (nanoparticles,\cite{PhysRevB.73.085412} periodic arrays,\cite{PhysRevB.71.035424} etc.) has attracted considerable attention in the past decade due to various intriguing applications ranging from spaser\cite{PhysRevLett.90.027402} and plasmonic nanolasers\cite{Noginov:2009aa,doi:10.1021/nl303086r} through the control of spontaneous emission\cite{Ropp:2013aa} to single-photon optical elements\cite{Chang:2007aa}. 

In the strong coupling regime, when the coupling strength of emitters and corresponding plasmon modes significantly exceeds any intrinsic decay rates, the hybrid states are formed. Depending on the coupling such states have both plasmonic and emitters properties allowing one to efficiently control the energy distribution in these systems in both linear\cite{doi:10.1021/nn101973p} and nonlinear\cite{doi:10.1021/nn4054528} regimes. The physics here is similar to the vacuum Rabi splitting phenomenon observed in systems composed of resonant atoms and high-Q macrocavities.\cite{PhysRevLett.68.1132} Furthermore the strong coupling mechanism between light and quantum emitters has been intensively investigated in semiconductor materials with semiconductor quantum wells positioned in microcavities such as high-Q Bragg mirrors.\cite{RevModPhys.71.1591} The exciton-normal mode coupling in semiconductor macrocavities leads to two transmission and reflection peaks/dips. The energy separation between these extrema called Rabi splitting (RS) is determined by the coupling strength between normal modes of a cavity and the excitonic transition in a quantum well. In such systems RS values were reported as high as $0.8$ meV.\cite{Khitrova:2006aa} This translates to a characteristic time of the energy exchange between coupled modes of about $5$ ps. 

One of the major advantages of plasmon sustaining materials over microcavities are much shorter time scales. The highest experimentally observed Rabi splitting reported up-to-date is $700$ meV\cite{PhysRevLett.106.196405} resulting in possible control of EM energy on a femtosecond timescale.\cite{Vasa:2013aa,doi:10.1021/nn4054528} Due to a high controllability of plasmonic materials one can experimentally tune optical  resonances of a given system to almost any imaginable energy in the visible with the great flexibility\cite{doi:10.1063/1.1578518} allowing not only to control EM radiation in the nanoscale\cite{PhysRevLett.95.093901,doi:10.1021/nl0524896} but also enabling spatial control of quantum systems.\cite{Aeschlimann:2007aa} 

To be able to control optical properties of materials in a spatiotemporal manner one may use the coherent properties of both laser radiation and surface plasmon-polaritons (SPPs). The original idea to extend coherent control techniques developed in atomic and molecular physics to plasmonics was proposed in Ref. [\onlinecite{PhysRevLett.88.067402}]. It was shown that it is possible to concentrate EM energy at a given point near a nanosystem using specific phase modulation of the incident radiation. This idea was further extended and extensively scrutinized by several other groups\cite{PhysRevB.71.035423,0953-4075-40-11-S04,0022-3727-41-19-195102,doi:10.1021/nl101285t,PhysRevB.83.205425,Yannopapas2011196,PhysRevApplied.1.014007} More specifically related to the present study are works discussing ultrafast control of spatial EM energy localization using linear chirps\cite{PhysRevB.71.035423,doi:10.1021/nl101285t}. As it was shown given a plasmonic system that supports several distinct SPP resonances one can use particular phase modulated incident radiation to control which part of the system is excited. The origin of such a control relies on the fact the different SPP resonances are associated with distinct spatial distributions of the EM energy in the near field.

In the present work I extend the ideas of spatiotemporal control in plasmonics to hybrid systems. By applying positive and negative linear chirps to sinusoidal plasmonic periodic gratings covered by a thin layer of molecules I show that transmission properties may be significantly altered. Its is shown that non-symmetric chirped pulses can be effectively used to control quantum emitters coupled to SPP modes in the strong coupling regime. The paper is organized as follows. The model and its numerical implementation are discussed in details in the Section \ref{sec:model}. The discussion of results is presented in the Section \ref{sec:results}. The work is concluded in the Section \ref{sec:conclusion}.

\section{Model and numerical implementation}
\label{sec:model}
To simulate optical response of non-magnetic plasmonic nanostructures I solve the system of Maxwell's equations in the time domain for the electric, $\vec{E}$, and magnetic, $\vec{H}$, fields
\begin{subequations}
 \begin{eqnarray}
\mu_0\frac{\partial\vec{H}}{\partial t}&=&- \nabla \times\vec{E}, \label{Faraday}\\
\varepsilon_0\frac{\partial\vec{E}}{\partial t}&=& \nabla \times\vec{H}-\vec{J}, \label{Ampere}
 \end{eqnarray}
\end{subequations}
where the current density is denoted as $\vec{J}$, $\mu_0$ and $\varepsilon_0$ are the magnetic permeability and dielectric permittivity of vacuum, respectively. The dynamical response of the metal is simulated using the Drude model
\begin{equation}
 \label{Drude}
\varepsilon\left ( \omega \right)=\varepsilon\left (\infty\right)-\frac{\Omega_P^2}{\omega^2+i\Gamma\omega},
\end{equation}
here $\varepsilon\left (\infty\right)$ is the high frequency limit of the dielectric constant, $\Omega_P$ is the bulk plasma frequency, and $\Gamma$ is the phenomenological damping. In simulations presented in the paper I consider silver described by the following numerical parameters: $\varepsilon\left (\infty\right)=8.926$, $\Omega_P = 11.585$ eV, $\Gamma=0.203$ eV.\cite{PhysRevB.68.045415} The linear response of conductive electrons results in a simple equation governing the dynamics of the current density\cite{Judkins:95}
\begin{equation}
 \label{density_current}
\frac{\partial\vec{J}}{\partial t}=\alpha\vec{J}+\beta\vec{E},
\end{equation}
with parameters $\alpha=-\Gamma$ and $\beta=\varepsilon_0\Omega_P^2$.

The system of equations (\ref{Faraday}), (\ref{Ampere}), (\ref{density_current}) is discretized in space and time and integrated using finite-difference time-domain (FDTD) technique.\cite{taflove2000computational} As an example of a plasmonic system I consider a one-dimensional periodic grating with a sinusoidal modulation of the silver film (see the inset of Fig. \ref{figure1}). Upper and lower boundaries of the grid are terminated with perfectly matched layers absorbing boundaries.\cite{berenger2007perfectly} Simple periodic boundary conditions are applied to EM fields at the right and left boundaries. To ensure numerical convergence all simulations including hybrid materials are performed with a spatial resolution of $1$ nm and temporal step of $0.0017$ fs. The system is excited with a plane wave using the total field/scattered field formalism at the normal incidence.\cite{taflove2000computational}

The optics of quantum emitters is scrutinized via self-consistent solution of Maxwell-Bloch equations, where the current density, $\vec{J}$, in the eq. (\ref{Ampere}) is replaced by the macroscopic polarization density, $\vec{P}$. The dynamics of the latter is simulated in accordance with the corresponding Liouville-von Neumann equation
\begin{equation}
 \label{Liouville}
i\hbar\frac{d\hat{\rho}}{d t}=\left[\hat{H},\hat{\rho}\right]-i\hbar\hat{\Gamma}(\hat{\rho}).
\end{equation}
In this work two level emitters are considered. Due to strong scattering of EM radiation that occurs at plasmon resonances all possible polarizations of the local electric field should be included in the total Hamiltonian $\hat{H}$. Both the Hamiltonian and the damping operator $\hat{\Gamma}$ are taken in the form discussed in details in Ref. [\onlinecite{PhysRevA.84.043802}]. In the two-dimensional geometry considered here the excited state of each emitter is a doubly degenerate state to account for $x$ and $y$ polarizations of $\vec{E}$.

From the density matrix, $\hat{\rho}$, the macroscopic polarization is evaluated according to 
\begin{equation}
 \label{polarization}
\vec{P}=n_0\text{Tr}\left(\hat{\rho}\vec{d} \right),
\end{equation}
where $n_0$ is the number density of emitters and $\vec{d}$ is the operator of emitter's dipole moment. To account for dipole-dipole interactions within a single grid cell in the FDTD numerical domain the local electric field, $\vec{E}_{\text{local}}$, that determines the quantum dynamics of emitters is modified to include local microscopic polarization also known as Lorentz-Lorenz correction\cite{PhysRevA.47.1247}
\begin{equation}
 \label{Lorentz-Lorenz}
\vec{E}_{\text{local}}=\vec{E}+\frac{\vec{P}}{3\varepsilon_0}.
\end{equation}

The numerical integration of the coupled Maxwell-Bloch equations (\ref{Faraday}), (\ref{Ampere}), (\ref{Liouville}) coupled via (\ref{polarization}) is performed using {\it weakly coupled} method, where EM field and density matrix are split in time by a half a time step and propagated in accordance with the leapfrog time stepping technique.\cite{NUM:NUM10046}

\section{Results and discussion}
\label{sec:results}

\begin{figure}[t!]
\begin{center}
\includegraphics[width=0.48\textwidth]{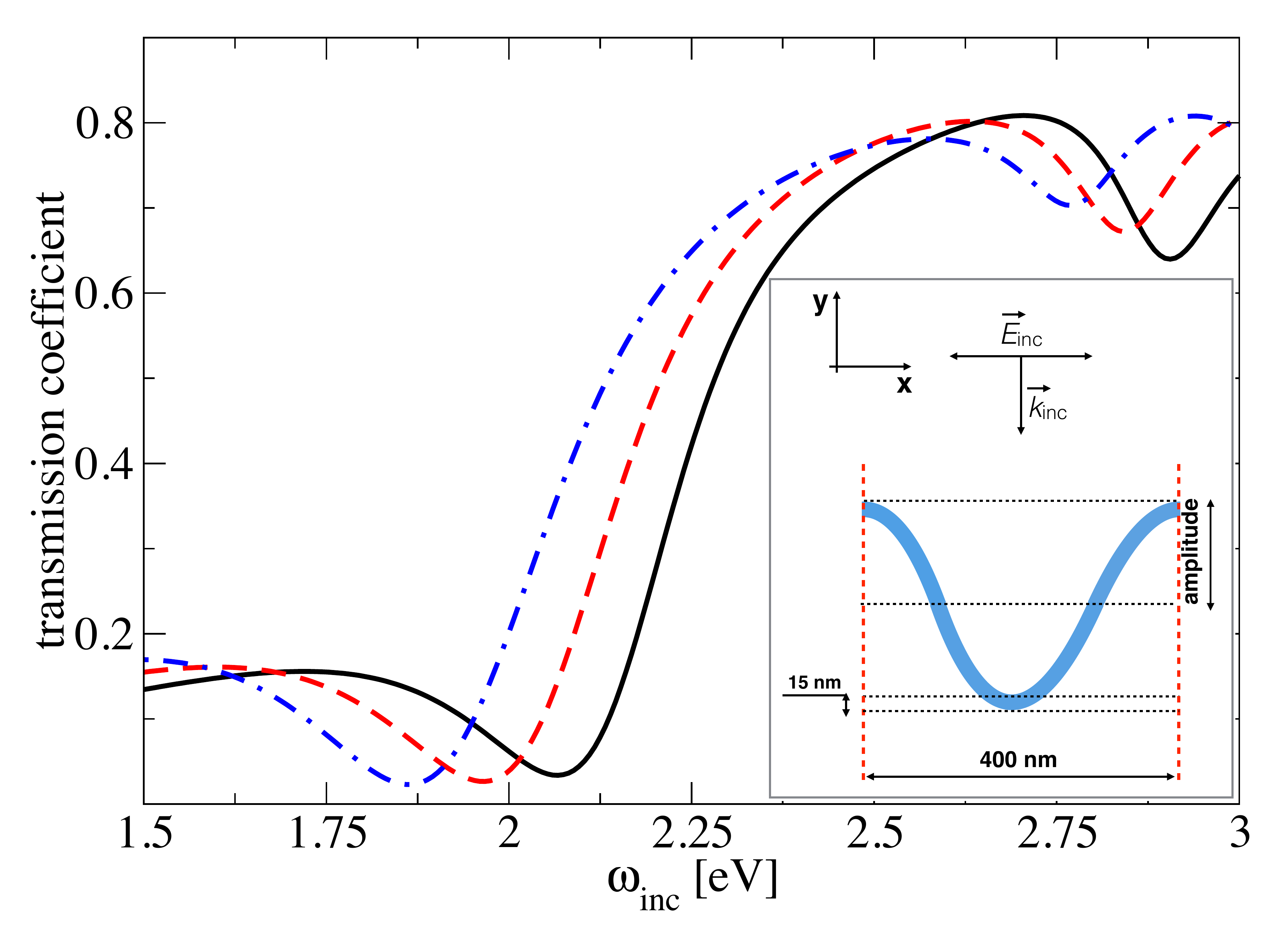}
\caption{\label{figure1} (Color online) Linear transmission coefficient as a function of the incident photon energy (in eV) calculated for a sinusoidal grating schematically depicted in the inset. The black solid line shows results for the grating with a modulation amplitude of $50$ nm, the red dashed line is for the amplitude of $60$ nm, and the blue dash-dotted line presents data for the amplitude of $70$ nm. Other parameters are shown in the inset.}
\end{center}
\end{figure}
The incident radiation is coupled to surface waves via diffraction grating schematically depicted in the inset of Fig. \ref{figure1}. If the thickness of the grating is on the oder of the skin depth for silver (about $20$ nm) evanescent EM modes on the input and output sides of the grating are strongly interacting. In the symmetric dielectric environment this results in formation of two distinct SPP modes: long- and short-lived SPP modes.\cite{doi:10.1063/1.3177011}
As a thickness of the film becomes smaller the interaction of the surface modes increases giving rise to SPP resonances with very high Qs\cite{PhysRev.182.539} and ultimately leading to very high local EM fields.\cite{doi:10.1063/1.3177011} For the purpose of this work it is vital to realize that such a system is highly tunable with SPP resonances dependent on both thickness of the film and its amplitude modulation as illustrated in the main panel of Fig. \ref{figure1}. I also note that although it is quite challenging to realize such a system experimentally\cite{1367-2630-10-6-065017} its close analog in symmetric dielectric environment has been made.\cite{Mu:10}

The linear transmission coefficient is evaluated via numerical integration of the outgoing EM energy flux on the output side of the grating along a detection line in the far-field zone (in the presented simulations this was approximately $1.1$ $\mu$m away from the film's surface). The integrated flux is then normalized with respect to the incident flux. The main panel of Fig. \ref{figure1} shows transmission as a function of the incident photon energy at three modulation amplitudes. The system exhibits two clear SPP resonances with energies significantly dependent upon the coupling of incident radiation to a given SPP mode. It is important to note that the corresponding reflection coefficient (not shown) exhibits maxima at the energies of the transmission minima confirming that the observed resonances are indeed surface modes. 

The corresponding spatial distributions of two EM modes significantly differ from each other suggesting an exciting opportunity to use SPP waves as a nanoscale probe to manipulate quantum emitters in the vicinity of the grating. As an example of a hybrid system I consider a sinusoidal plasmonic grating covered by a $20$ nm thin molecular layer that has two distinct absorption resonances. By properly adjusting the modulation amplitude one may tune SPP modes of a bare grating to molecular transition energies. Note that the resonant energies of SPP modes do not scale identically as one varies the modulation amplitude as seen from Fig. \ref{figure1}.

\begin{figure}[t!]
\begin{center}
\includegraphics[width=0.48\textwidth]{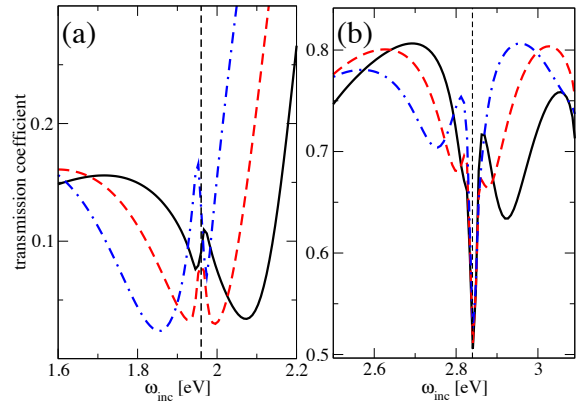}
\caption{\label{figure2} (Color online) Both panels show linear transmission coefficient as a function of the incident photon energy (in eV) calculated for a hybrid system comprised of a sinusoidal silver grating and a 20 nm thin molecular layer deposited on the input side of the grating. Panel (a) shows transmission near the low energy SPP mode. Panel (b) presents data near the high energy SPP resonance. Black solid lines show data for the grating with a modulation amplitude of 50 nm. Red dashed lines present results obtained for the grating with an amplitude of 60 nm. Blue dash-dotted lines show results for the amplitude of 70 nm. The molecular layer exhibits two distinct absorption resonances, energies of which are indicated in each panel by vertical dashed lines: $1.96$ eV (type I) and $2.84$ eV (type II). Panel (a) presents data near the low energy SPP mode. Panel (b) shows results for the high energy SPP mode. Other parameters of simulations are: the thickness of the silver film is $15$ nm, the period of the grating is $400$ nm, the molecular transition dipole for both transitions is $25$ Debye, the number density of molecules with each transition is $n_0=5\times10^{24}$ m$/$s$^2$, the pure dephasing time for both molecular transitions is $100$ fs, the radiationless lifetime of molecular excited states at both transitions is $1$ ps.}
\end{center}
\end{figure}

Fig. \ref{figure2} shows the linear transmission spectra for the sinusoidal grating covered with a 20 nm thin resonant molecular layer. The molecular layer has two separate distinct absorption lines (two types of emitters) at $1.96$ eV and $2.84$ eV referred further in the text as type I and type II molecules, respectively . Several interesting features are worth noting: 
\begin{description}
  \item[(a)] when the lower energy SPP mode in the Fig. \ref{figure2}a sweeps through the molecular line at $1.96$ eV (indicated as a vertical dashed line) one can clearly see a splitting and formation of two hybrid modes - a unique feature of the strong coupling between SPPs and type I emitters. The energy splitting between hybrid modes varies from $70$ meV for the grating with the modulation amplitude of $60$ nm to $130$ meV seen for the modulation amplitude of $50$ nm;
  \item[(b)]  in the Fig. \ref{figure2}b the interaction of the type II emitters with the high energy SPP mode is more complex compared to the lower energy mode as the former exhibits an extra-mode at the exact resonance (i.e. when the emitters' transition energy matches that of the SPP mode) for the grating with the modulation amplitude of $60$ nm. The presence of the third mode in an otherwise two-mode spectrum in hybrid systems has been noted several times and recently was explained in Ref. [\onlinecite{PhysRevLett.109.073002}] as a signature of the collective long-range emitter-emitter interaction strongly enhanced by plasmons. Even though the number density of emitters in Fig. \ref{figure2} is not that high (there is no third mode observed for lower energy SPPs) the local EM field associated with the high energy SPPs is significantly higher.
\end{description}

\begin{figure}[t!]
\begin{center}
\includegraphics[width=0.48\textwidth]{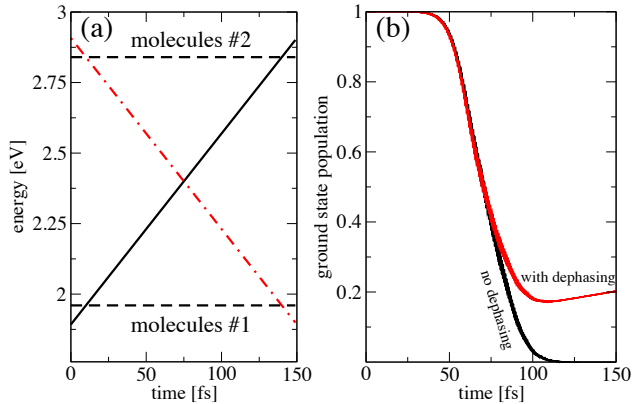}
\caption{\label{figure3} (Color online) Illustration of the non-symmetric chirp control. Panel (a) shows energies of two types of molecules (horizontal dashed lines) used in simulations and the incident laser frequency for positive  (solid black line) and negative (red dash-dotted line) chirps as functions of time. Panel (b) presents results of simulations for a single molecule interacting with a non-symmetric chirp. Here the ground state population is shown as a function of time. Two curves are obtained for a molecule without dephasing and relaxation damping (black line) and with dephasing and relaxation taken into account (red line). Simulations are carried out for a positive chirp and type II molecules (shown as upper horizontal dashed line in panel (a)). The rest of the molecular parameters are the same as in previous Figure. The incident laser pulse is $150$ fs long with a peak amplitude of $10^9$ V$/$m.}
\end{center}
\end{figure}

Since the local EM fields associated with SPP modes of the grating have significantly different spatial distribution one may employ spatiotemporal control of transmission in such a hybrid system by tuning incident radiation into either of the SPP resonances thus controlling not only which emitters are excited but also where on the surface they are excited. To accomplish this one may employ adiabatic passage technique developed in atomic\cite{doi:10.1146/annurev.physchem.52.1.763} and molecular physics\cite{PhysRevA.63.043415}. This however would require relatively long pulses. In order to operate on a femtosecond timescale it is proposed to use linearly chirped laser pulses that are offset in time. The idea is illustrated in Fig. \ref{figure3}a. Here horizontal dashed lines represent energies of two types of molecules deposited onto a plasmonic grating. The incident pulse duration is set at $150$ fs with a frequency changing in time as shown in the Fig. \ref{figure3}a. Consider a positive chirp, for instance: type I molecules will barely get excited since the incident frequency passes the resonance at the beginning of the pulse while the type II molecules will undergo considerable excitation as the incident frequency approaches the resonance at the end of the pulse. 

The incident pulse used in simulations is
\begin{equation}
 \label{incident_pulse}
\vec{E}_{\text{inc}}=\vec{E}_0f\left(t\right)\text{cos}\left(\omega\left(t\right)\right),
\end{equation}
where the time envelope $f\left(t\right)$ is taken in the form of the Blackman-Harris window function.\cite{MOP:MOP14} The time dependence of the incident frequency is parameterized according to
\begin{subequations}
 \begin{eqnarray}
\omega\left(t \right)=\omega_0+g\times\frac{t}{\tau}, \label{omega_t}\\
\omega_0=\frac{\omega_1\tau-\delta\tau\left( \omega_1+\omega_2\right)}{\tau-\delta\tau}, \label{omega0}\\
g=\frac{\tau\left( \omega_2-\omega_1\right)}{\tau-\delta\tau}, \label{alpha_chirp}
 \end{eqnarray}
\end{subequations}
where $\tau$ is the pulse duration, $\delta\tau$ is the time at which the incident frequency reaches one of the molecular resonances, i.e. $\omega\left(\delta\tau \right)=\omega_1$ and $\omega\left(\tau-\delta\tau \right)=\omega_2$.

\begin{figure}[t!]
\begin{center}
\includegraphics[width=0.48\textwidth]{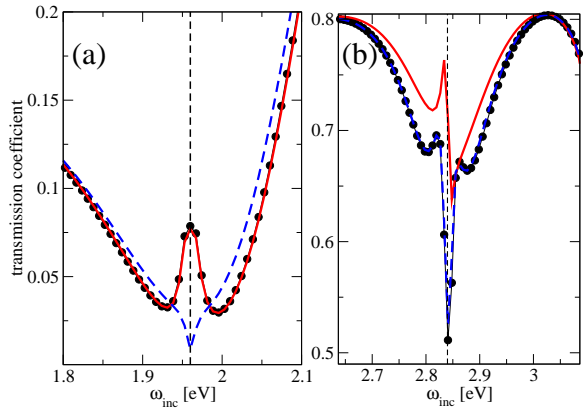}
\caption{\label{figure4} (Color online) Controlling transmission with non-symmetric chirps. Both panels show linear transmission coefficient as a function of the incident photon energy (in eV) calculated for a hybrid system comprised of a sinusoidal silver grating with the modulation amplitude of $60$ nm and a $20$ nm thin molecular layer deposited on the input side of the grating with the period of $400$ nm. Panel (a) shows transmission near the low energy SPP mode. Panel (b) presents data near the high energy SPP resonance. The molecular layer contains two types of molecules with transition energies of $1.96$ eV and $2.84$ eV as in Fig. \ref{figure2}. The transition energies are shown as vertical dashed lines. Black circles show transmission before a chirped pulse is applied. Solid red line shows transmission after the positive chirp excites the hybrid material (see Fig. \ref{figure3}a, solid black line). Blue dashed line shows transmission after the negative chirp is applied (see Fig. \ref{figure3}a, dashed red line). The molecular parameters of simulations are the same as in Fig. \ref{figure2}.  The incident laser pulse is $150$ fs long with a peak amplitude of $10^9$ V$/$m.}
\end{center}
\end{figure}

Fig. \ref{figure3}b shows the results of simulations performed for a single molecule case without plasmonic material present. Simulations are carried out for type II molecules with the incident field having a positive chirp, frequency of which is plotted in Fig. \ref{figure3}a (black solid line). The numerical values of the chirp parameters (\ref{omega_t}) are: $\tau=150$ fs, $\delta\tau=10$ fs, $\omega_1=1.96$ eV, and $\omega_2=2.84$ eV. The ground state population of molecules smoothly changes from $1$ to $10^{-6}$ for the case without damping and to $0.2$ with the dephasing and relaxation parameters taken into account. Simulations performed for the same chirp but for the type I molecules show nearly no disturbance in the population. Similarly if one changes the sign of the chirp (red dashed line in Fig. \ref{figure3}a) and propagates density matrix for the first type of molecules the dynamics will be nearly identical to the one shown in Fig. \ref{figure3}b with no change in populations for the type II of molecules. It is hence possible to controllably manipulate a given type of quantum emitters by changing the sign of the linear chirp. Moreover this would allow to alter significantly optical properties of a hybrid system at a given frequency while keeping intact its response in the other parts of the spectrum. It should be noted that even though the local EM field enhancement is anticipated in hybrid systems supporting plasmon resonances it should not worsen the proposed scheme but rather improve it. Simulations for a single molecule case were also performed at high incident field amplitudes confirming the robustness of the scheme. A possible problem, however, may arise from the fact that plasmons not only enhance the local EM field but also alter its spectrum due to material dispersion. If the local field increases and is no longer a simple linear chirp the scheme would most likely not work.

\begin{figure}[t!]
\begin{center}
\includegraphics[width=0.48\textwidth]{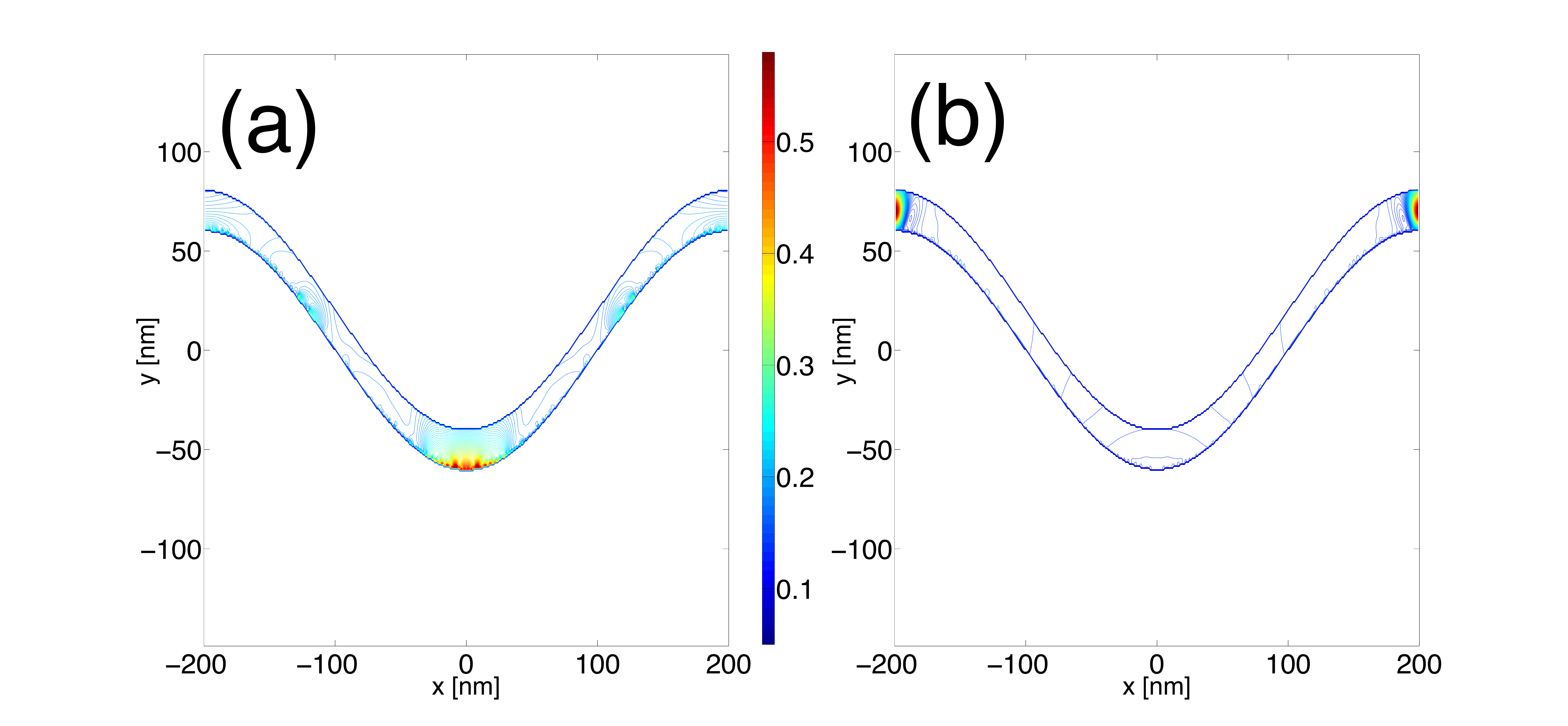}
\caption{\label{figure5} (Color online) Spatial distributions of the molecular ground state population in the molecular layer after a chirped pulse excitation. Panel (a) shows the distribution of the ground state population of the molecules with the transition energy of $2.84$ eV after the excitation with a positive chirp (see Fig. \ref{figure3}a, solid black line). Panel (b) shows the ground state population for the molecules with the energy $1.96$ eV after being excited with a negative chirp (see Fig. \ref{figure3}a, dashed red line). Parameters of simulations are the same as in Fig. \ref{figure4}.}
\end{center}
\end{figure}

The main results of this paper are shown in Fig. \ref{figure4}. Positive and negative chirps are applied to excite a sinusoidal grating with a molecular layer containing two types of molecules (resonant at $1.96$ eV and $2.84$ eV). After the system is excited and the incident pulse is turned off the linear transmission is calculated. The general idea of using non-symmetric chirps to alter different molecules independently and hence manipulate transmission properties at different frequencies is clearly illustrated. The negative chirp inverts the type I molecules significantly changing transmission (Fig. \ref{figure4}a) while molecules resonant at $2.84$ eV are not perturbed at all (Fig. \ref{figure4}b) and vice versa. It is however important to point out that the effect of the positive chirp is not as dramatic as the negative chirp. Upon careful examination of the dynamics it is possible to conclude that the positive chirp does not act as a $\pi$-pulse, i.e. does not gradually invert the corresponding molecules. The molecules rather undergo several Rabi cycles. A plausible explanation is that the high energy SPP mode contributes additional frequency components to the local EM field.

It is also informative to examine spatial distributions of the molecular populations after a chirped pulse excitation. This is shown in Fig. \ref{figure5} for two cases considered here. It is seen that positive and negative chirps not only excite different types of molecules as anticipated but do so in a spatially different manner. The high energy molecules are partially in the ground state in the middle of the molecular layer with others all inverted. The positive chirp inverts low energy molecules everywhere in the layer except for a small region at the top.

\section{Conclusion}
\label{sec:conclusion}
Using self-consistent model of coupled Maxwell-Bloch equations I examined the interaction of linear chirped laser pulses with a hybrid system comprised of a sinusoidal diffraction grating and a molecular layer. It is shown that one may use non-symmetric chirps to control which molecular transitions are excited. The proposed scheme is scrutinized in case of two types of molecules resonantly coupled to two separate SPP modes. It is also demonstrated that the control may also be achieved is a spatial manner due to highly inhomogeneous localization of EM energy at plasmon resonances.

\bibliography{references_library}

\begin{thebibliography}{48}%
\makeatletter
\providecommand \@ifxundefined [1]{%
 \@ifx{#1\undefined}
}%
\providecommand \@ifnum [1]{%
 \ifnum #1\expandafter \@firstoftwo
 \else \expandafter \@secondoftwo
 \fi
}%
\providecommand \@ifx [1]{%
 \ifx #1\expandafter \@firstoftwo
 \else \expandafter \@secondoftwo
 \fi
}%
\providecommand \natexlab [1]{#1}%
\providecommand \enquote  [1]{``#1''}%
\providecommand \bibnamefont  [1]{#1}%
\providecommand \bibfnamefont [1]{#1}%
\providecommand \citenamefont [1]{#1}%
\providecommand \href@noop [0]{\@secondoftwo}%
\providecommand \href [0]{\begingroup \@sanitize@url \@href}%
\providecommand \@href[1]{\@@startlink{#1}\@@href}%
\providecommand \@@href[1]{\endgroup#1\@@endlink}%
\providecommand \@sanitize@url [0]{\catcode `\\12\catcode `\$12\catcode
  `\&12\catcode `\#12\catcode `\^12\catcode `\_12\catcode `\%12\relax}%
\providecommand \@@startlink[1]{}%
\providecommand \@@endlink[0]{}%
\providecommand \url  [0]{\begingroup\@sanitize@url \@url }%
\providecommand \@url [1]{\endgroup\@href {#1}{\urlprefix }}%
\providecommand \urlprefix  [0]{URL }%
\providecommand \Eprint [0]{\href }%
\providecommand \doibase [0]{http://dx.doi.org/}%
\providecommand \selectlanguage [0]{\@gobble}%
\providecommand \bibinfo  [0]{\@secondoftwo}%
\providecommand \bibfield  [0]{\@secondoftwo}%
\providecommand \translation [1]{[#1]}%
\providecommand \BibitemOpen [0]{}%
\providecommand \bibitemStop [0]{}%
\providecommand \bibitemNoStop [0]{.\EOS\space}%
\providecommand \EOS [0]{\spacefactor3000\relax}%
\providecommand \BibitemShut  [1]{\csname bibitem#1\endcsname}%
\let\auto@bib@innerbib\@empty
\bibitem [{\citenamefont {Murray}\ and\ \citenamefont
  {Barnes}(2007)}]{ADMA:ADMA200700678}%
  \BibitemOpen
  \bibfield  {author} {\bibinfo {author} {\bibfnamefont {W.~A.}\ \bibnamefont
  {Murray}}\ and\ \bibinfo {author} {\bibfnamefont {W.~L.}\ \bibnamefont
  {Barnes}},\ }\bibfield  {title} {\enquote {\bibinfo {title} {Plasmonic
  materials},}\ }\href {\doibase 10.1002/adma.200700678} {\bibfield  {journal}
  {\bibinfo  {journal} {Adv. Mater.}\ }\textbf {\bibinfo {volume} {19}},\
  \bibinfo {pages} {3771--3782} (\bibinfo {year} {2007})}\BibitemShut {NoStop}%
\bibitem [{\citenamefont {Gramotnev}\ and\ \citenamefont
  {Bozhevolnyi}(2010)}]{Gramotnev:2010aa}%
  \BibitemOpen
  \bibfield  {author} {\bibinfo {author} {\bibfnamefont {D.~K.}\ \bibnamefont
  {Gramotnev}}\ and\ \bibinfo {author} {\bibfnamefont {S.~I.}\ \bibnamefont
  {Bozhevolnyi}},\ }\bibfield  {title} {\enquote {\bibinfo {title} {Plasmonics
  beyond the diffraction limit},}\ }\href
  {http://dx.doi.org/10.1038/nphoton.2009.282} {\bibfield  {journal} {\bibinfo
  {journal} {Nat. Photon.}\ }\textbf {\bibinfo {volume} {4}},\ \bibinfo {pages}
  {83--91} (\bibinfo {year} {2010})}\BibitemShut {NoStop}%
\bibitem [{\citenamefont {Schuller}\ \emph {et~al.}(2010)\citenamefont
  {Schuller}, \citenamefont {Barnard}, \citenamefont {Cai}, \citenamefont
  {Jun}, \citenamefont {White},\ and\ \citenamefont
  {Brongersma}}]{Schuller:2010aa}%
  \BibitemOpen
  \bibfield  {author} {\bibinfo {author} {\bibfnamefont {J.~A.}\ \bibnamefont
  {Schuller}}, \bibinfo {author} {\bibfnamefont {E.~S.}\ \bibnamefont
  {Barnard}}, \bibinfo {author} {\bibfnamefont {W.}~\bibnamefont {Cai}},
  \bibinfo {author} {\bibfnamefont {Y.~C.}\ \bibnamefont {Jun}}, \bibinfo
  {author} {\bibfnamefont {J.~S.}\ \bibnamefont {White}}, \ and\ \bibinfo
  {author} {\bibfnamefont {M.~L.}\ \bibnamefont {Brongersma}},\ }\bibfield
  {title} {\enquote {\bibinfo {title} {Plasmonics for extreme light
  concentration and manipulation},}\ }\href
  {http://dx.doi.org/10.1038/nmat2630} {\bibfield  {journal} {\bibinfo
  {journal} {Nat. Mater.}\ }\textbf {\bibinfo {volume} {9}},\ \bibinfo {pages}
  {193--204} (\bibinfo {year} {2010})}\BibitemShut {NoStop}%
\bibitem [{\citenamefont {Stockman}(2011)}]{Stockman:11}%
  \BibitemOpen
  \bibfield  {author} {\bibinfo {author} {\bibfnamefont {M.~I.}\ \bibnamefont
  {Stockman}},\ }\bibfield  {title} {\enquote {\bibinfo {title}
  {Nanoplasmonics: past, present, and glimpse into future},}\ }\href {\doibase
  10.1364/OE.19.022029} {\bibfield  {journal} {\bibinfo  {journal} {Opt.
  Express}\ }\textbf {\bibinfo {volume} {19}},\ \bibinfo {pages} {22029--22106}
  (\bibinfo {year} {2011})}\BibitemShut {NoStop}%
\bibitem [{\citenamefont {T{\"o}rm{\"a}}\ and\ \citenamefont
  {Barnes}(2014)}]{torma2014strong}%
  \BibitemOpen
  \bibfield  {author} {\bibinfo {author} {\bibfnamefont {P.}~\bibnamefont
  {T{\"o}rm{\"a}}}\ and\ \bibinfo {author} {\bibfnamefont {W.}~\bibnamefont
  {Barnes}},\ }\bibfield  {title} {\enquote {\bibinfo {title} {Strong coupling
  between surface plasmon polaritons and emitters},}\ }\href@noop {} {\bibfield
   {journal} {\bibinfo  {journal} {arXiv preprint arXiv:1405.1661}\ } (\bibinfo
  {year} {2014})}\BibitemShut {NoStop}%
\bibitem [{\citenamefont {Fedutik}\ \emph {et~al.}(2007)\citenamefont
  {Fedutik}, \citenamefont {Temnov}, \citenamefont {Sch\"ops}, \citenamefont
  {Woggon},\ and\ \citenamefont {Artemyev}}]{PhysRevLett.99.136802}%
  \BibitemOpen
  \bibfield  {author} {\bibinfo {author} {\bibfnamefont {Y.}~\bibnamefont
  {Fedutik}}, \bibinfo {author} {\bibfnamefont {V.~V.}\ \bibnamefont {Temnov}},
  \bibinfo {author} {\bibfnamefont {O.}~\bibnamefont {Sch\"ops}}, \bibinfo
  {author} {\bibfnamefont {U.}~\bibnamefont {Woggon}}, \ and\ \bibinfo {author}
  {\bibfnamefont {M.~V.}\ \bibnamefont {Artemyev}},\ }\bibfield  {title}
  {\enquote {\bibinfo {title} {Exciton-plasmon-photon conversion in plasmonic
  nanostructures},}\ }\href {\doibase 10.1103/PhysRevLett.99.136802} {\bibfield
   {journal} {\bibinfo  {journal} {Phys. Rev. Lett.}\ }\textbf {\bibinfo
  {volume} {99}},\ \bibinfo {pages} {136802} (\bibinfo {year}
  {2007})}\BibitemShut {NoStop}%
\bibitem [{\citenamefont {Bellessa}\ \emph {et~al.}(2004)\citenamefont
  {Bellessa}, \citenamefont {Bonnand}, \citenamefont {Plenet},\ and\
  \citenamefont {Mugnier}}]{PhysRevLett.93.036404}%
  \BibitemOpen
  \bibfield  {author} {\bibinfo {author} {\bibfnamefont {J.}~\bibnamefont
  {Bellessa}}, \bibinfo {author} {\bibfnamefont {C.}~\bibnamefont {Bonnand}},
  \bibinfo {author} {\bibfnamefont {J.~C.}\ \bibnamefont {Plenet}}, \ and\
  \bibinfo {author} {\bibfnamefont {J.}~\bibnamefont {Mugnier}},\ }\bibfield
  {title} {\enquote {\bibinfo {title} {Strong coupling between surface plasmons
  and excitons in an organic semiconductor},}\ }\href {\doibase
  10.1103/PhysRevLett.93.036404} {\bibfield  {journal} {\bibinfo  {journal}
  {Phys. Rev. Lett.}\ }\textbf {\bibinfo {volume} {93}},\ \bibinfo {pages}
  {036404} (\bibinfo {year} {2004})}\BibitemShut {NoStop}%
\bibitem [{\citenamefont {Ambj\"ornsson}\ \emph {et~al.}(2006)\citenamefont
  {Ambj\"ornsson}, \citenamefont {Mukhopadhyay}, \citenamefont {Apell},\ and\
  \citenamefont {K\"all}}]{PhysRevB.73.085412}%
  \BibitemOpen
  \bibfield  {author} {\bibinfo {author} {\bibfnamefont {T.}~\bibnamefont
  {Ambj\"ornsson}}, \bibinfo {author} {\bibfnamefont {G.}~\bibnamefont
  {Mukhopadhyay}}, \bibinfo {author} {\bibfnamefont {S.~P.}\ \bibnamefont
  {Apell}}, \ and\ \bibinfo {author} {\bibfnamefont {M.}~\bibnamefont
  {K\"all}},\ }\bibfield  {title} {\enquote {\bibinfo {title} {Resonant
  coupling between localized plasmons and anisotropic molecular coatings in
  ellipsoidal metal nanoparticles},}\ }\href {\doibase
  10.1103/PhysRevB.73.085412} {\bibfield  {journal} {\bibinfo  {journal} {Phys.
  Rev. B}\ }\textbf {\bibinfo {volume} {73}},\ \bibinfo {pages} {085412}
  (\bibinfo {year} {2006})}\BibitemShut {NoStop}%
\bibitem [{\citenamefont {Dintinger}\ \emph {et~al.}(2005)\citenamefont
  {Dintinger}, \citenamefont {Klein}, \citenamefont {Bustos}, \citenamefont
  {Barnes},\ and\ \citenamefont {Ebbesen}}]{PhysRevB.71.035424}%
  \BibitemOpen
  \bibfield  {author} {\bibinfo {author} {\bibfnamefont {J.}~\bibnamefont
  {Dintinger}}, \bibinfo {author} {\bibfnamefont {S.}~\bibnamefont {Klein}},
  \bibinfo {author} {\bibfnamefont {F.}~\bibnamefont {Bustos}}, \bibinfo
  {author} {\bibfnamefont {W.~L.}\ \bibnamefont {Barnes}}, \ and\ \bibinfo
  {author} {\bibfnamefont {T.~W.}\ \bibnamefont {Ebbesen}},\ }\bibfield
  {title} {\enquote {\bibinfo {title} {Strong coupling between surface
  plasmon-polaritons and organic molecules in subwavelength hole arrays},}\
  }\href {\doibase 10.1103/PhysRevB.71.035424} {\bibfield  {journal} {\bibinfo
  {journal} {Phys. Rev. B}\ }\textbf {\bibinfo {volume} {71}},\ \bibinfo
  {pages} {035424} (\bibinfo {year} {2005})}\BibitemShut {NoStop}%
\bibitem [{\citenamefont {Bergman}\ and\ \citenamefont
  {Stockman}(2003)}]{PhysRevLett.90.027402}%
  \BibitemOpen
  \bibfield  {author} {\bibinfo {author} {\bibfnamefont {D.~J.}\ \bibnamefont
  {Bergman}}\ and\ \bibinfo {author} {\bibfnamefont {M.~I.}\ \bibnamefont
  {Stockman}},\ }\bibfield  {title} {\enquote {\bibinfo {title} {Surface
  plasmon amplification by stimulated emission of radiation: Quantum generation
  of coherent surface plasmons in nanosystems},}\ }\href {\doibase
  10.1103/PhysRevLett.90.027402} {\bibfield  {journal} {\bibinfo  {journal}
  {Phys. Rev. Lett.}\ }\textbf {\bibinfo {volume} {90}},\ \bibinfo {pages}
  {027402} (\bibinfo {year} {2003})}\BibitemShut {NoStop}%
\bibitem [{\citenamefont {Noginov}\ \emph {et~al.}(2009)\citenamefont
  {Noginov}, \citenamefont {Zhu}, \citenamefont {Belgrave}, \citenamefont
  {Bakker}, \citenamefont {Shalaev}, \citenamefont {Narimanov}, \citenamefont
  {Stout}, \citenamefont {Herz}, \citenamefont {Suteewong},\ and\ \citenamefont
  {Wiesner}}]{Noginov:2009aa}%
  \BibitemOpen
  \bibfield  {author} {\bibinfo {author} {\bibfnamefont {M.~A.}\ \bibnamefont
  {Noginov}}, \bibinfo {author} {\bibfnamefont {G.}~\bibnamefont {Zhu}},
  \bibinfo {author} {\bibfnamefont {A.~M.}\ \bibnamefont {Belgrave}}, \bibinfo
  {author} {\bibfnamefont {R.}~\bibnamefont {Bakker}}, \bibinfo {author}
  {\bibfnamefont {V.~M.}\ \bibnamefont {Shalaev}}, \bibinfo {author}
  {\bibfnamefont {E.~E.}\ \bibnamefont {Narimanov}}, \bibinfo {author}
  {\bibfnamefont {S.}~\bibnamefont {Stout}}, \bibinfo {author} {\bibfnamefont
  {E.}~\bibnamefont {Herz}}, \bibinfo {author} {\bibfnamefont {T.}~\bibnamefont
  {Suteewong}}, \ and\ \bibinfo {author} {\bibfnamefont {U.}~\bibnamefont
  {Wiesner}},\ }\bibfield  {title} {\enquote {\bibinfo {title} {Demonstration
  of a spaser-based nanolaser},}\ }\href
  {http://dx.doi.org/10.1038/nature08318} {\bibfield  {journal} {\bibinfo
  {journal} {Nature}\ }\textbf {\bibinfo {volume} {460}},\ \bibinfo {pages}
  {1110--1112} (\bibinfo {year} {2009})}\BibitemShut {NoStop}%
\bibitem [{\citenamefont {Suh}\ \emph {et~al.}(2012)\citenamefont {Suh},
  \citenamefont {Kim}, \citenamefont {Zhou}, \citenamefont {Huntington},
  \citenamefont {Co}, \citenamefont {Wasielewski},\ and\ \citenamefont
  {Odom}}]{doi:10.1021/nl303086r}%
  \BibitemOpen
  \bibfield  {author} {\bibinfo {author} {\bibfnamefont {J.~Y.}\ \bibnamefont
  {Suh}}, \bibinfo {author} {\bibfnamefont {C.~H.}\ \bibnamefont {Kim}},
  \bibinfo {author} {\bibfnamefont {W.}~\bibnamefont {Zhou}}, \bibinfo {author}
  {\bibfnamefont {M.~D.}\ \bibnamefont {Huntington}}, \bibinfo {author}
  {\bibfnamefont {D.~T.}\ \bibnamefont {Co}}, \bibinfo {author} {\bibfnamefont
  {M.~R.}\ \bibnamefont {Wasielewski}}, \ and\ \bibinfo {author} {\bibfnamefont
  {T.~W.}\ \bibnamefont {Odom}},\ }\bibfield  {title} {\enquote {\bibinfo
  {title} {Plasmonic bowtie nanolaser arrays},}\ }\href {\doibase
  10.1021/nl303086r} {\bibfield  {journal} {\bibinfo  {journal} {Nano Lett.}\
  }\textbf {\bibinfo {volume} {12}},\ \bibinfo {pages} {5769--5774} (\bibinfo
  {year} {2012})}\BibitemShut {NoStop}%
\bibitem [{\citenamefont {Ropp}\ \emph {et~al.}(2013)\citenamefont {Ropp},
  \citenamefont {Cummins}, \citenamefont {Nah}, \citenamefont {Fourkas},
  \citenamefont {Shapiro},\ and\ \citenamefont {Waks}}]{Ropp:2013aa}%
  \BibitemOpen
  \bibfield  {author} {\bibinfo {author} {\bibfnamefont {C.}~\bibnamefont
  {Ropp}}, \bibinfo {author} {\bibfnamefont {Z.}~\bibnamefont {Cummins}},
  \bibinfo {author} {\bibfnamefont {S.}~\bibnamefont {Nah}}, \bibinfo {author}
  {\bibfnamefont {J.~T.}\ \bibnamefont {Fourkas}}, \bibinfo {author}
  {\bibfnamefont {B.}~\bibnamefont {Shapiro}}, \ and\ \bibinfo {author}
  {\bibfnamefont {E.}~\bibnamefont {Waks}},\ }\bibfield  {title} {\enquote
  {\bibinfo {title} {Nanoscale imaging and spontaneous emission control with a
  single nano-positioned quantum dot},}\ }\href
  {http://dx.doi.org/10.1038/ncomms2477} {\bibfield  {journal} {\bibinfo
  {journal} {Nat. Commun.}\ }\textbf {\bibinfo {volume} {4}},\ \bibinfo {pages}
  {1447} (\bibinfo {year} {2013})}\BibitemShut {NoStop}%
\bibitem [{\citenamefont {Chang}\ \emph {et~al.}(2007)\citenamefont {Chang},
  \citenamefont {Sorensen}, \citenamefont {Demler},\ and\ \citenamefont
  {Lukin}}]{Chang:2007aa}%
  \BibitemOpen
  \bibfield  {author} {\bibinfo {author} {\bibfnamefont {D.~E.}\ \bibnamefont
  {Chang}}, \bibinfo {author} {\bibfnamefont {A.~S.}\ \bibnamefont {Sorensen}},
  \bibinfo {author} {\bibfnamefont {E.~A.}\ \bibnamefont {Demler}}, \ and\
  \bibinfo {author} {\bibfnamefont {M.~D.}\ \bibnamefont {Lukin}},\ }\bibfield
  {title} {\enquote {\bibinfo {title} {A single-photon transistor using
  nanoscale surface plasmons},}\ }\href {http://dx.doi.org/10.1038/nphys708}
  {\bibfield  {journal} {\bibinfo  {journal} {Nat. Phys.}\ }\textbf {\bibinfo
  {volume} {3}},\ \bibinfo {pages} {807--812} (\bibinfo {year}
  {2007})}\BibitemShut {NoStop}%
\bibitem [{\citenamefont {Vasa}\ \emph {et~al.}(2010)\citenamefont {Vasa},
  \citenamefont {Pomraenke}, \citenamefont {Cirmi}, \citenamefont {De~Re},
  \citenamefont {Wang}, \citenamefont {Schwieger}, \citenamefont {Leipold},
  \citenamefont {Runge}, \citenamefont {Cerullo},\ and\ \citenamefont
  {Lienau}}]{doi:10.1021/nn101973p}%
  \BibitemOpen
  \bibfield  {author} {\bibinfo {author} {\bibfnamefont {P.}~\bibnamefont
  {Vasa}}, \bibinfo {author} {\bibfnamefont {R.}~\bibnamefont {Pomraenke}},
  \bibinfo {author} {\bibfnamefont {G.}~\bibnamefont {Cirmi}}, \bibinfo
  {author} {\bibfnamefont {E.}~\bibnamefont {De~Re}}, \bibinfo {author}
  {\bibfnamefont {W.}~\bibnamefont {Wang}}, \bibinfo {author} {\bibfnamefont
  {S.}~\bibnamefont {Schwieger}}, \bibinfo {author} {\bibfnamefont
  {D.}~\bibnamefont {Leipold}}, \bibinfo {author} {\bibfnamefont
  {E.}~\bibnamefont {Runge}}, \bibinfo {author} {\bibfnamefont
  {G.}~\bibnamefont {Cerullo}}, \ and\ \bibinfo {author} {\bibfnamefont
  {C.}~\bibnamefont {Lienau}},\ }\bibfield  {title} {\enquote {\bibinfo {title}
  {Ultrafast energy transfer between molecular assemblies and surface plasmons
  in the strong coupling regime},}\ }\href {\doibase 10.1021/nn101973p}
  {\bibfield  {journal} {\bibinfo  {journal} {ACS Nano}\ }\textbf {\bibinfo
  {volume} {4}},\ \bibinfo {pages} {7559--7565} (\bibinfo {year}
  {2010})}\BibitemShut {NoStop}%
\bibitem [{\citenamefont {Sukharev}\ \emph {et~al.}(2014)\citenamefont
  {Sukharev}, \citenamefont {Seideman}, \citenamefont {Gordon}, \citenamefont
  {Salomon},\ and\ \citenamefont {Prior}}]{doi:10.1021/nn4054528}%
  \BibitemOpen
  \bibfield  {author} {\bibinfo {author} {\bibfnamefont {M.}~\bibnamefont
  {Sukharev}}, \bibinfo {author} {\bibfnamefont {T.}~\bibnamefont {Seideman}},
  \bibinfo {author} {\bibfnamefont {R.~J.}\ \bibnamefont {Gordon}}, \bibinfo
  {author} {\bibfnamefont {A.}~\bibnamefont {Salomon}}, \ and\ \bibinfo
  {author} {\bibfnamefont {Y.}~\bibnamefont {Prior}},\ }\bibfield  {title}
  {\enquote {\bibinfo {title} {Ultrafast energy transfer between molecular
  assemblies and surface plasmons in the strong coupling regime},}\ }\href
  {\doibase 10.1021/nn4054528} {\bibfield  {journal} {\bibinfo  {journal} {ACS
  Nano}\ }\textbf {\bibinfo {volume} {8}},\ \bibinfo {pages} {807--817}
  (\bibinfo {year} {2014})}\BibitemShut {NoStop}%
\bibitem [{\citenamefont {Thompson}, \citenamefont {Rempe},\ and\ \citenamefont
  {Kimble}(1992)}]{PhysRevLett.68.1132}%
  \BibitemOpen
  \bibfield  {author} {\bibinfo {author} {\bibfnamefont {R.~J.}\ \bibnamefont
  {Thompson}}, \bibinfo {author} {\bibfnamefont {G.}~\bibnamefont {Rempe}}, \
  and\ \bibinfo {author} {\bibfnamefont {H.~J.}\ \bibnamefont {Kimble}},\
  }\bibfield  {title} {\enquote {\bibinfo {title} {Observation of normal-mode
  splitting for an atom in an optical cavity},}\ }\href {\doibase
  10.1103/PhysRevLett.68.1132} {\bibfield  {journal} {\bibinfo  {journal}
  {Phys. Rev. Lett.}\ }\textbf {\bibinfo {volume} {68}},\ \bibinfo {pages}
  {1132--1135} (\bibinfo {year} {1992})}\BibitemShut {NoStop}%
\bibitem [{\citenamefont {Khitrova}\ \emph {et~al.}(1999)\citenamefont
  {Khitrova}, \citenamefont {Gibbs}, \citenamefont {Jahnke}, \citenamefont
  {Kira},\ and\ \citenamefont {Koch}}]{RevModPhys.71.1591}%
  \BibitemOpen
  \bibfield  {author} {\bibinfo {author} {\bibfnamefont {G.}~\bibnamefont
  {Khitrova}}, \bibinfo {author} {\bibfnamefont {H.~M.}\ \bibnamefont {Gibbs}},
  \bibinfo {author} {\bibfnamefont {F.}~\bibnamefont {Jahnke}}, \bibinfo
  {author} {\bibfnamefont {M.}~\bibnamefont {Kira}}, \ and\ \bibinfo {author}
  {\bibfnamefont {S.~W.}\ \bibnamefont {Koch}},\ }\bibfield  {title} {\enquote
  {\bibinfo {title} {Nonlinear optics of normal-mode-coupling semiconductor
  microcavities},}\ }\href {\doibase 10.1103/RevModPhys.71.1591} {\bibfield
  {journal} {\bibinfo  {journal} {Rev. Mod. Phys.}\ }\textbf {\bibinfo {volume}
  {71}},\ \bibinfo {pages} {1591--1639} (\bibinfo {year} {1999})}\BibitemShut
  {NoStop}%
\bibitem [{\citenamefont {Khitrova}\ \emph {et~al.}(2006)\citenamefont
  {Khitrova}, \citenamefont {Gibbs}, \citenamefont {Kira}, \citenamefont
  {Koch},\ and\ \citenamefont {Scherer}}]{Khitrova:2006aa}%
  \BibitemOpen
  \bibfield  {author} {\bibinfo {author} {\bibfnamefont {G.}~\bibnamefont
  {Khitrova}}, \bibinfo {author} {\bibfnamefont {H.~M.}\ \bibnamefont {Gibbs}},
  \bibinfo {author} {\bibfnamefont {M.}~\bibnamefont {Kira}}, \bibinfo {author}
  {\bibfnamefont {S.~W.}\ \bibnamefont {Koch}}, \ and\ \bibinfo {author}
  {\bibfnamefont {A.}~\bibnamefont {Scherer}},\ }\bibfield  {title} {\enquote
  {\bibinfo {title} {Vacuum rabi splitting in semiconductors},}\ }\href
  {http://dx.doi.org/10.1038/nphys227} {\bibfield  {journal} {\bibinfo
  {journal} {Nat. Phys.}\ }\textbf {\bibinfo {volume} {2}},\ \bibinfo {pages}
  {81--90} (\bibinfo {year} {2006})}\BibitemShut {NoStop}%
\bibitem [{\citenamefont {Schwartz}\ \emph {et~al.}(2011)\citenamefont
  {Schwartz}, \citenamefont {Hutchison}, \citenamefont {Genet},\ and\
  \citenamefont {Ebbesen}}]{PhysRevLett.106.196405}%
  \BibitemOpen
  \bibfield  {author} {\bibinfo {author} {\bibfnamefont {T.}~\bibnamefont
  {Schwartz}}, \bibinfo {author} {\bibfnamefont {J.~A.}\ \bibnamefont
  {Hutchison}}, \bibinfo {author} {\bibfnamefont {C.}~\bibnamefont {Genet}}, \
  and\ \bibinfo {author} {\bibfnamefont {T.~W.}\ \bibnamefont {Ebbesen}},\
  }\bibfield  {title} {\enquote {\bibinfo {title} {Reversible switching of
  ultrastrong light-molecule coupling},}\ }\href {\doibase
  10.1103/PhysRevLett.106.196405} {\bibfield  {journal} {\bibinfo  {journal}
  {Phys. Rev. Lett.}\ }\textbf {\bibinfo {volume} {106}},\ \bibinfo {pages}
  {196405} (\bibinfo {year} {2011})}\BibitemShut {NoStop}%
\bibitem [{\citenamefont {Vasa}\ \emph {et~al.}(2013)\citenamefont {Vasa},
  \citenamefont {Wang}, \citenamefont {Pomraenke}, \citenamefont {Lammers},
  \citenamefont {Maiuri}, \citenamefont {Manzoni}, \citenamefont {Cerullo},\
  and\ \citenamefont {Lienau}}]{Vasa:2013aa}%
  \BibitemOpen
  \bibfield  {author} {\bibinfo {author} {\bibfnamefont {P.}~\bibnamefont
  {Vasa}}, \bibinfo {author} {\bibfnamefont {W.}~\bibnamefont {Wang}}, \bibinfo
  {author} {\bibfnamefont {R.}~\bibnamefont {Pomraenke}}, \bibinfo {author}
  {\bibfnamefont {M.}~\bibnamefont {Lammers}}, \bibinfo {author} {\bibfnamefont
  {M.}~\bibnamefont {Maiuri}}, \bibinfo {author} {\bibfnamefont
  {C.}~\bibnamefont {Manzoni}}, \bibinfo {author} {\bibfnamefont
  {G.}~\bibnamefont {Cerullo}}, \ and\ \bibinfo {author} {\bibfnamefont
  {C.}~\bibnamefont {Lienau}},\ }\bibfield  {title} {\enquote {\bibinfo {title}
  {Real-time observation of ultrafast rabi oscillations between excitons and
  plasmons in metal nanostructures with j-aggregates},}\ }\href
  {http://dx.doi.org/10.1038/nphoton.2012.340} {\bibfield  {journal} {\bibinfo
  {journal} {Nat. Photon.}\ }\textbf {\bibinfo {volume} {7}},\ \bibinfo {pages}
  {128--132} (\bibinfo {year} {2013})}\BibitemShut {NoStop}%
\bibitem [{\citenamefont {Xu}\ \emph {et~al.}(2003)\citenamefont {Xu},
  \citenamefont {Tazawa}, \citenamefont {Jin}, \citenamefont {Nakao},\ and\
  \citenamefont {Yoshimura}}]{doi:10.1063/1.1578518}%
  \BibitemOpen
  \bibfield  {author} {\bibinfo {author} {\bibfnamefont {G.}~\bibnamefont
  {Xu}}, \bibinfo {author} {\bibfnamefont {M.}~\bibnamefont {Tazawa}}, \bibinfo
  {author} {\bibfnamefont {P.}~\bibnamefont {Jin}}, \bibinfo {author}
  {\bibfnamefont {S.}~\bibnamefont {Nakao}}, \ and\ \bibinfo {author}
  {\bibfnamefont {K.}~\bibnamefont {Yoshimura}},\ }\bibfield  {title} {\enquote
  {\bibinfo {title} {Wavelength tuning of surface plasmon resonance using
  dielectric layers on silver island films},}\ }\href {\doibase
  http://dx.doi.org/10.1063/1.1578518} {\bibfield  {journal} {\bibinfo
  {journal} {Appl. Phys. Lett.}\ }\textbf {\bibinfo {volume} {82}},\ \bibinfo
  {pages} {3811--3813} (\bibinfo {year} {2003})}\BibitemShut {NoStop}%
\bibitem [{\citenamefont {Brixner}\ \emph {et~al.}(2005)\citenamefont
  {Brixner}, \citenamefont {Garcia~de Abajo}, \citenamefont {Schneider},\ and\
  \citenamefont {Pfeiffer}}]{PhysRevLett.95.093901}%
  \BibitemOpen
  \bibfield  {author} {\bibinfo {author} {\bibfnamefont {T.}~\bibnamefont
  {Brixner}}, \bibinfo {author} {\bibfnamefont {F.~J.}\ \bibnamefont {Garcia~de
  Abajo}}, \bibinfo {author} {\bibfnamefont {J.}~\bibnamefont {Schneider}}, \
  and\ \bibinfo {author} {\bibfnamefont {W.}~\bibnamefont {Pfeiffer}},\
  }\bibfield  {title} {\enquote {\bibinfo {title} {Nanoscopic ultrafast
  space-time-resolved spectroscopy},}\ }\href {\doibase
  10.1103/PhysRevLett.95.093901} {\bibfield  {journal} {\bibinfo  {journal}
  {Phys. Rev. Lett.}\ }\textbf {\bibinfo {volume} {95}},\ \bibinfo {pages}
  {093901} (\bibinfo {year} {2005})}\BibitemShut {NoStop}%
\bibitem [{\citenamefont {Sukharev}\ and\ \citenamefont
  {Seideman}(2006)}]{doi:10.1021/nl0524896}%
  \BibitemOpen
  \bibfield  {author} {\bibinfo {author} {\bibfnamefont {M.}~\bibnamefont
  {Sukharev}}\ and\ \bibinfo {author} {\bibfnamefont {T.}~\bibnamefont
  {Seideman}},\ }\bibfield  {title} {\enquote {\bibinfo {title} {Phase and
  polarization control as a route to plasmonic nanodevices},}\ }\href {\doibase
  10.1021/nl0524896} {\bibfield  {journal} {\bibinfo  {journal} {Nano Lett.}\
  }\textbf {\bibinfo {volume} {6}},\ \bibinfo {pages} {715--719} (\bibinfo
  {year} {2006})}\BibitemShut {NoStop}%
\bibitem [{\citenamefont {Aeschlimann}\ \emph {et~al.}(2007)\citenamefont
  {Aeschlimann}, \citenamefont {Bauer}, \citenamefont {Bayer}, \citenamefont
  {Brixner}, \citenamefont {Garcia~de Abajo}, \citenamefont {Pfeiffer},
  \citenamefont {Rohmer}, \citenamefont {Spindler},\ and\ \citenamefont
  {Steeb}}]{Aeschlimann:2007aa}%
  \BibitemOpen
  \bibfield  {author} {\bibinfo {author} {\bibfnamefont {M.}~\bibnamefont
  {Aeschlimann}}, \bibinfo {author} {\bibfnamefont {M.}~\bibnamefont {Bauer}},
  \bibinfo {author} {\bibfnamefont {D.}~\bibnamefont {Bayer}}, \bibinfo
  {author} {\bibfnamefont {T.}~\bibnamefont {Brixner}}, \bibinfo {author}
  {\bibfnamefont {F.~J.}\ \bibnamefont {Garcia~de Abajo}}, \bibinfo {author}
  {\bibfnamefont {W.}~\bibnamefont {Pfeiffer}}, \bibinfo {author}
  {\bibfnamefont {M.}~\bibnamefont {Rohmer}}, \bibinfo {author} {\bibfnamefont
  {C.}~\bibnamefont {Spindler}}, \ and\ \bibinfo {author} {\bibfnamefont
  {F.}~\bibnamefont {Steeb}},\ }\bibfield  {title} {\enquote {\bibinfo {title}
  {Adaptive subwavelength control of nano-optical fields},}\ }\href
  {http://dx.doi.org/10.1038/nature05595} {\bibfield  {journal} {\bibinfo
  {journal} {Nature}\ }\textbf {\bibinfo {volume} {446}},\ \bibinfo {pages}
  {301--304} (\bibinfo {year} {2007})}\BibitemShut {NoStop}%
\bibitem [{\citenamefont {Stockman}, \citenamefont {Faleev},\ and\
  \citenamefont {Bergman}(2002)}]{PhysRevLett.88.067402}%
  \BibitemOpen
  \bibfield  {author} {\bibinfo {author} {\bibfnamefont {M.~I.}\ \bibnamefont
  {Stockman}}, \bibinfo {author} {\bibfnamefont {S.~V.}\ \bibnamefont
  {Faleev}}, \ and\ \bibinfo {author} {\bibfnamefont {D.~J.}\ \bibnamefont
  {Bergman}},\ }\bibfield  {title} {\enquote {\bibinfo {title} {Coherent
  control of femtosecond energy localization in nanosystems},}\ }\href
  {\doibase 10.1103/PhysRevLett.88.067402} {\bibfield  {journal} {\bibinfo
  {journal} {Phys. Rev. Lett.}\ }\textbf {\bibinfo {volume} {88}},\ \bibinfo
  {pages} {067402} (\bibinfo {year} {2002})}\BibitemShut {NoStop}%
\bibitem [{\citenamefont {Lee}\ and\ \citenamefont
  {Gray}(2005)}]{PhysRevB.71.035423}%
  \BibitemOpen
  \bibfield  {author} {\bibinfo {author} {\bibfnamefont {T.-W.}\ \bibnamefont
  {Lee}}\ and\ \bibinfo {author} {\bibfnamefont {S.~K.}\ \bibnamefont {Gray}},\
  }\bibfield  {title} {\enquote {\bibinfo {title} {Controlled spatiotemporal
  excitation of metal nanoparticles with picosecond optical pulses},}\ }\href
  {\doibase 10.1103/PhysRevB.71.035423} {\bibfield  {journal} {\bibinfo
  {journal} {Phys. Rev. B}\ }\textbf {\bibinfo {volume} {71}},\ \bibinfo
  {pages} {035423} (\bibinfo {year} {2005})}\BibitemShut {NoStop}%
\bibitem [{\citenamefont {Sukharev}\ and\ \citenamefont
  {Seideman}(2007)}]{0953-4075-40-11-S04}%
  \BibitemOpen
  \bibfield  {author} {\bibinfo {author} {\bibfnamefont {M.}~\bibnamefont
  {Sukharev}}\ and\ \bibinfo {author} {\bibfnamefont {T.}~\bibnamefont
  {Seideman}},\ }\bibfield  {title} {\enquote {\bibinfo {title} {Coherent
  control of light propagation via nanoparticle arrays},}\ }\href
  {http://stacks.iop.org/0953-4075/40/i=11/a=S04} {\bibfield  {journal}
  {\bibinfo  {journal} {J. Phys. B - At. Mol. Opt.}\ }\textbf {\bibinfo
  {volume} {40}},\ \bibinfo {pages} {S283} (\bibinfo {year}
  {2007})}\BibitemShut {NoStop}%
\bibitem [{\citenamefont {Guyader}\ \emph {et~al.}(2008)\citenamefont
  {Guyader}, \citenamefont {Kirilyuk}, \citenamefont {Rasing}, \citenamefont
  {Wurtz}, \citenamefont {Zayats}, \citenamefont {Alkemade},\ and\
  \citenamefont {Smolyaninov}}]{0022-3727-41-19-195102}%
  \BibitemOpen
  \bibfield  {author} {\bibinfo {author} {\bibfnamefont {L.~L.}\ \bibnamefont
  {Guyader}}, \bibinfo {author} {\bibfnamefont {A.}~\bibnamefont {Kirilyuk}},
  \bibinfo {author} {\bibfnamefont {T.}~\bibnamefont {Rasing}}, \bibinfo
  {author} {\bibfnamefont {G.~A.}\ \bibnamefont {Wurtz}}, \bibinfo {author}
  {\bibfnamefont {A.~V.}\ \bibnamefont {Zayats}}, \bibinfo {author}
  {\bibfnamefont {P.~F.~A.}\ \bibnamefont {Alkemade}}, \ and\ \bibinfo {author}
  {\bibfnamefont {I.~I.}\ \bibnamefont {Smolyaninov}},\ }\bibfield  {title}
  {\enquote {\bibinfo {title} {Coherent control of surface plasmon polariton
  mediated optical transmission},}\ }\href
  {http://stacks.iop.org/0022-3727/41/i=19/a=195102} {\bibfield  {journal}
  {\bibinfo  {journal} {J. Phys. D Appl. Phys.}\ }\textbf {\bibinfo {volume}
  {41}},\ \bibinfo {pages} {195102} (\bibinfo {year} {2008})}\BibitemShut
  {NoStop}%
\bibitem [{\citenamefont {Cao}\ \emph {et~al.}(2010)\citenamefont {Cao},
  \citenamefont {Nome}, \citenamefont {Montgomery}, \citenamefont {Gray},\ and\
  \citenamefont {Scherer}}]{doi:10.1021/nl101285t}%
  \BibitemOpen
  \bibfield  {author} {\bibinfo {author} {\bibfnamefont {L.}~\bibnamefont
  {Cao}}, \bibinfo {author} {\bibfnamefont {R.~A.}\ \bibnamefont {Nome}},
  \bibinfo {author} {\bibfnamefont {J.~M.}\ \bibnamefont {Montgomery}},
  \bibinfo {author} {\bibfnamefont {S.~K.}\ \bibnamefont {Gray}}, \ and\
  \bibinfo {author} {\bibfnamefont {N.~F.}\ \bibnamefont {Scherer}},\
  }\bibfield  {title} {\enquote {\bibinfo {title} {Controlling plasmonic wave
  packets in silver nanowires},}\ }\href {\doibase 10.1021/nl101285t}
  {\bibfield  {journal} {\bibinfo  {journal} {Nano Lett.}\ }\textbf {\bibinfo
  {volume} {10}},\ \bibinfo {pages} {3389--3394} (\bibinfo {year}
  {2010})}\BibitemShut {NoStop}%
\bibitem [{\citenamefont {Fainberg}\ \emph {et~al.}(2011)\citenamefont
  {Fainberg}, \citenamefont {Sukharev}, \citenamefont {Park},\ and\
  \citenamefont {Galperin}}]{PhysRevB.83.205425}%
  \BibitemOpen
  \bibfield  {author} {\bibinfo {author} {\bibfnamefont {B.~D.}\ \bibnamefont
  {Fainberg}}, \bibinfo {author} {\bibfnamefont {M.}~\bibnamefont {Sukharev}},
  \bibinfo {author} {\bibfnamefont {T.-H.}\ \bibnamefont {Park}}, \ and\
  \bibinfo {author} {\bibfnamefont {M.}~\bibnamefont {Galperin}},\ }\bibfield
  {title} {\enquote {\bibinfo {title} {Light-induced current in molecular
  junctions: Local field and non-markov effects},}\ }\href {\doibase
  10.1103/PhysRevB.83.205425} {\bibfield  {journal} {\bibinfo  {journal} {Phys.
  Rev. B}\ }\textbf {\bibinfo {volume} {83}},\ \bibinfo {pages} {205425}
  (\bibinfo {year} {2011})}\BibitemShut {NoStop}%
\bibitem [{\citenamefont {Yannopapas}\ and\ \citenamefont
  {Vitanov}(2011)}]{Yannopapas2011196}%
  \BibitemOpen
  \bibfield  {author} {\bibinfo {author} {\bibfnamefont {V.}~\bibnamefont
  {Yannopapas}}\ and\ \bibinfo {author} {\bibfnamefont {N.~V.}\ \bibnamefont
  {Vitanov}},\ }\bibfield  {title} {\enquote {\bibinfo {title} {Coherent
  control of surface exciton-polaritons in collections of semiconductor
  nanoparticles: A theoretical study},}\ }\href {\doibase
  http://dx.doi.org/10.1016/j.photonics.2010.07.003} {\bibfield  {journal}
  {\bibinfo  {journal} {Photonics and Nanostructures - Fundamentals and
  Applications}\ }\textbf {\bibinfo {volume} {9}},\ \bibinfo {pages} {196 --
  200} (\bibinfo {year} {2011})},\ \bibinfo {note} {emerging Trends and Novel
  Materials in Photonics}\BibitemShut {NoStop}%
\bibitem [{\citenamefont {Rewitz}\ \emph {et~al.}(2014)\citenamefont {Rewitz},
  \citenamefont {Razinskas}, \citenamefont {Geisler}, \citenamefont {Krauss},
  \citenamefont {Goetz}, \citenamefont {Paw\l{}owska}, \citenamefont {Hecht},\
  and\ \citenamefont {Brixner}}]{PhysRevApplied.1.014007}%
  \BibitemOpen
  \bibfield  {author} {\bibinfo {author} {\bibfnamefont {C.}~\bibnamefont
  {Rewitz}}, \bibinfo {author} {\bibfnamefont {G.}~\bibnamefont {Razinskas}},
  \bibinfo {author} {\bibfnamefont {P.}~\bibnamefont {Geisler}}, \bibinfo
  {author} {\bibfnamefont {E.}~\bibnamefont {Krauss}}, \bibinfo {author}
  {\bibfnamefont {S.}~\bibnamefont {Goetz}}, \bibinfo {author} {\bibfnamefont
  {M.}~\bibnamefont {Paw\l{}owska}}, \bibinfo {author} {\bibfnamefont
  {B.}~\bibnamefont {Hecht}}, \ and\ \bibinfo {author} {\bibfnamefont
  {T.}~\bibnamefont {Brixner}},\ }\bibfield  {title} {\enquote {\bibinfo
  {title} {Coherent control of plasmon propagation in a nanocircuit},}\ }\href
  {\doibase 10.1103/PhysRevApplied.1.014007} {\bibfield  {journal} {\bibinfo
  {journal} {Phys. Rev. Applied}\ }\textbf {\bibinfo {volume} {1}},\ \bibinfo
  {pages} {014007} (\bibinfo {year} {2014})}\BibitemShut {NoStop}%
\bibitem [{\citenamefont {Gray}\ and\ \citenamefont
  {Kupka}(2003)}]{PhysRevB.68.045415}%
  \BibitemOpen
  \bibfield  {author} {\bibinfo {author} {\bibfnamefont {S.~K.}\ \bibnamefont
  {Gray}}\ and\ \bibinfo {author} {\bibfnamefont {T.}~\bibnamefont {Kupka}},\
  }\bibfield  {title} {\enquote {\bibinfo {title} {Propagation of light in
  metallic nanowire arrays:\quad{}finite-difference time-domain studies of
  silver cylinders},}\ }\href {\doibase 10.1103/PhysRevB.68.045415} {\bibfield
  {journal} {\bibinfo  {journal} {Phys. Rev. B}\ }\textbf {\bibinfo {volume}
  {68}},\ \bibinfo {pages} {045415} (\bibinfo {year} {2003})}\BibitemShut
  {NoStop}%
\bibitem [{\citenamefont {Judkins}\ and\ \citenamefont
  {Ziolkowski}(1995)}]{Judkins:95}%
  \BibitemOpen
  \bibfield  {author} {\bibinfo {author} {\bibfnamefont {J.~B.}\ \bibnamefont
  {Judkins}}\ and\ \bibinfo {author} {\bibfnamefont {R.~W.}\ \bibnamefont
  {Ziolkowski}},\ }\bibfield  {title} {\enquote {\bibinfo {title}
  {Finite-difference time-domain modeling of nonperfectly conducting metallic
  thin-film gratings},}\ }\href {\doibase 10.1364/JOSAA.12.001974} {\bibfield
  {journal} {\bibinfo  {journal} {J. Opt. Soc. Am. A}\ }\textbf {\bibinfo
  {volume} {12}},\ \bibinfo {pages} {1974--1983} (\bibinfo {year}
  {1995})}\BibitemShut {NoStop}%
\bibitem [{\citenamefont {Taflove}\ and\ \citenamefont
  {Hagness}(2000)}]{taflove2000computational}%
  \BibitemOpen
  \bibfield  {author} {\bibinfo {author} {\bibfnamefont {A.}~\bibnamefont
  {Taflove}}\ and\ \bibinfo {author} {\bibfnamefont {S.}~\bibnamefont
  {Hagness}},\ }\href@noop {} {\emph {\bibinfo {title} {Computational
  Electrodynamics: The Finite-Difference Time-Domain Method}}}\ (\bibinfo
  {publisher} {Artech House},\ \bibinfo {year} {2000})\BibitemShut {NoStop}%
\bibitem [{\citenamefont {B{\'e}renger}(2007)}]{berenger2007perfectly}%
  \BibitemOpen
  \bibfield  {author} {\bibinfo {author} {\bibfnamefont {J.-P.}\ \bibnamefont
  {B{\'e}renger}},\ }\bibfield  {title} {\enquote {\bibinfo {title} {Perfectly
  matched layer (pml) for computational electromagnetics},}\ }\href@noop {}
  {\bibfield  {journal} {\bibinfo  {journal} {Synthesis Lectures on
  Computational Electromagnetics}\ }\textbf {\bibinfo {volume} {2}},\ \bibinfo
  {pages} {1--117} (\bibinfo {year} {2007})}\BibitemShut {NoStop}%
\bibitem [{\citenamefont {Sukharev}\ and\ \citenamefont
  {Nitzan}(2011)}]{PhysRevA.84.043802}%
  \BibitemOpen
  \bibfield  {author} {\bibinfo {author} {\bibfnamefont {M.}~\bibnamefont
  {Sukharev}}\ and\ \bibinfo {author} {\bibfnamefont {A.}~\bibnamefont
  {Nitzan}},\ }\bibfield  {title} {\enquote {\bibinfo {title} {Numerical
  studies of the interaction of an atomic sample with the electromagnetic field
  in two dimensions},}\ }\href {\doibase 10.1103/PhysRevA.84.043802} {\bibfield
   {journal} {\bibinfo  {journal} {Phys. Rev. A}\ }\textbf {\bibinfo {volume}
  {84}},\ \bibinfo {pages} {043802} (\bibinfo {year} {2011})}\BibitemShut
  {NoStop}%
\bibitem [{\citenamefont {Bowden}\ and\ \citenamefont
  {Dowling}(1993)}]{PhysRevA.47.1247}%
  \BibitemOpen
  \bibfield  {author} {\bibinfo {author} {\bibfnamefont {C.~M.}\ \bibnamefont
  {Bowden}}\ and\ \bibinfo {author} {\bibfnamefont {J.~P.}\ \bibnamefont
  {Dowling}},\ }\bibfield  {title} {\enquote {\bibinfo {title}
  {Near\char21{}dipole-dipole effects in dense media: Generalized maxwell-bloch
  equations},}\ }\href {\doibase 10.1103/PhysRevA.47.1247} {\bibfield
  {journal} {\bibinfo  {journal} {Phys. Rev. A}\ }\textbf {\bibinfo {volume}
  {47}},\ \bibinfo {pages} {1247--1251} (\bibinfo {year} {1993})}\BibitemShut
  {NoStop}%
\bibitem [{\citenamefont {Bid{\'e}garay}(2003)}]{NUM:NUM10046}%
  \BibitemOpen
  \bibfield  {author} {\bibinfo {author} {\bibfnamefont {B.}~\bibnamefont
  {Bid{\'e}garay}},\ }\bibfield  {title} {\enquote {\bibinfo {title} {Time
  discretizations for maxwell-bloch equations},}\ }\href {\doibase
  10.1002/num.10046} {\bibfield  {journal} {\bibinfo  {journal} {Numerical
  Methods for Partial Differential Equations}\ }\textbf {\bibinfo {volume}
  {19}},\ \bibinfo {pages} {284--300} (\bibinfo {year} {2003})}\BibitemShut
  {NoStop}%
\bibitem [{\citenamefont {Sukharev}\ \emph {et~al.}(2009)\citenamefont
  {Sukharev}, \citenamefont {Sievert}, \citenamefont {Seideman},\ and\
  \citenamefont {Ketterson}}]{doi:10.1063/1.3177011}%
  \BibitemOpen
  \bibfield  {author} {\bibinfo {author} {\bibfnamefont {M.}~\bibnamefont
  {Sukharev}}, \bibinfo {author} {\bibfnamefont {P.~R.}\ \bibnamefont
  {Sievert}}, \bibinfo {author} {\bibfnamefont {T.}~\bibnamefont {Seideman}}, \
  and\ \bibinfo {author} {\bibfnamefont {J.~B.}\ \bibnamefont {Ketterson}},\
  }\bibfield  {title} {\enquote {\bibinfo {title} {Perfect coupling of light to
  surface plasmons with ultra-narrow linewidths},}\ }\href {\doibase
  http://dx.doi.org/10.1063/1.3177011} {\bibfield  {journal} {\bibinfo
  {journal} {J. Chem. Phys.}\ }\textbf {\bibinfo {volume} {131}},\ \bibinfo
  {eid} {034708} (\bibinfo {year} {2009})}\BibitemShut {NoStop}%
\bibitem [{\citenamefont {Economou}(1969)}]{PhysRev.182.539}%
  \BibitemOpen
  \bibfield  {author} {\bibinfo {author} {\bibfnamefont {E.~N.}\ \bibnamefont
  {Economou}},\ }\bibfield  {title} {\enquote {\bibinfo {title} {Surface
  plasmons in thin films},}\ }\href {\doibase 10.1103/PhysRev.182.539}
  {\bibfield  {journal} {\bibinfo  {journal} {Phys. Rev.}\ }\textbf {\bibinfo
  {volume} {182}},\ \bibinfo {pages} {539--554} (\bibinfo {year}
  {1969})}\BibitemShut {NoStop}%
\bibitem [{\citenamefont {Symonds}\ \emph {et~al.}(2008)\citenamefont
  {Symonds}, \citenamefont {Bonnand}, \citenamefont {Plenet}, \citenamefont
  {Br{\'e}hier}, \citenamefont {Parashkov}, \citenamefont {Lauret},
  \citenamefont {Deleporte},\ and\ \citenamefont
  {Bellessa}}]{1367-2630-10-6-065017}%
  \BibitemOpen
  \bibfield  {author} {\bibinfo {author} {\bibfnamefont {C.}~\bibnamefont
  {Symonds}}, \bibinfo {author} {\bibfnamefont {C.}~\bibnamefont {Bonnand}},
  \bibinfo {author} {\bibfnamefont {J.~C.}\ \bibnamefont {Plenet}}, \bibinfo
  {author} {\bibfnamefont {A.}~\bibnamefont {Br{\'e}hier}}, \bibinfo {author}
  {\bibfnamefont {R.}~\bibnamefont {Parashkov}}, \bibinfo {author}
  {\bibfnamefont {J.~S.}\ \bibnamefont {Lauret}}, \bibinfo {author}
  {\bibfnamefont {E.}~\bibnamefont {Deleporte}}, \ and\ \bibinfo {author}
  {\bibfnamefont {J.}~\bibnamefont {Bellessa}},\ }\bibfield  {title} {\enquote
  {\bibinfo {title} {Particularities of surface plasmon--exciton strong
  coupling with large rabi splitting},}\ }\href
  {http://stacks.iop.org/1367-2630/10/i=6/a=065017} {\bibfield  {journal}
  {\bibinfo  {journal} {New J. Phys.}\ }\textbf {\bibinfo {volume} {10}},\
  \bibinfo {pages} {065017} (\bibinfo {year} {2008})}\BibitemShut {NoStop}%
\bibitem [{\citenamefont {Mu}\ \emph {et~al.}(2010)\citenamefont {Mu},
  \citenamefont {Buchholz}, \citenamefont {Sukharev}, \citenamefont {Jang},
  \citenamefont {Chang},\ and\ \citenamefont {Ketterson}}]{Mu:10}%
  \BibitemOpen
  \bibfield  {author} {\bibinfo {author} {\bibfnamefont {W.}~\bibnamefont
  {Mu}}, \bibinfo {author} {\bibfnamefont {D.~B.}\ \bibnamefont {Buchholz}},
  \bibinfo {author} {\bibfnamefont {M.}~\bibnamefont {Sukharev}}, \bibinfo
  {author} {\bibfnamefont {J.~I.}\ \bibnamefont {Jang}}, \bibinfo {author}
  {\bibfnamefont {R.~P.}\ \bibnamefont {Chang}}, \ and\ \bibinfo {author}
  {\bibfnamefont {J.~B.}\ \bibnamefont {Ketterson}},\ }\bibfield  {title}
  {\enquote {\bibinfo {title} {One-dimensional long-range plasmonic-photonic
  structures},}\ }\href {\doibase 10.1364/OL.35.000550} {\bibfield  {journal}
  {\bibinfo  {journal} {Opt. Lett.}\ }\textbf {\bibinfo {volume} {35}},\
  \bibinfo {pages} {550--552} (\bibinfo {year} {2010})}\BibitemShut {NoStop}%
\bibitem [{\citenamefont {Salomon}\ \emph {et~al.}(2012)\citenamefont
  {Salomon}, \citenamefont {Gordon}, \citenamefont {Prior}, \citenamefont
  {Seideman},\ and\ \citenamefont {Sukharev}}]{PhysRevLett.109.073002}%
  \BibitemOpen
  \bibfield  {author} {\bibinfo {author} {\bibfnamefont {A.}~\bibnamefont
  {Salomon}}, \bibinfo {author} {\bibfnamefont {R.~J.}\ \bibnamefont {Gordon}},
  \bibinfo {author} {\bibfnamefont {Y.}~\bibnamefont {Prior}}, \bibinfo
  {author} {\bibfnamefont {T.}~\bibnamefont {Seideman}}, \ and\ \bibinfo
  {author} {\bibfnamefont {M.}~\bibnamefont {Sukharev}},\ }\bibfield  {title}
  {\enquote {\bibinfo {title} {Strong coupling between molecular excited states
  and surface plasmon modes of a slit array in a thin metal film},}\ }\href
  {\doibase 10.1103/PhysRevLett.109.073002} {\bibfield  {journal} {\bibinfo
  {journal} {Phys. Rev. Lett.}\ }\textbf {\bibinfo {volume} {109}},\ \bibinfo
  {pages} {073002} (\bibinfo {year} {2012})}\BibitemShut {NoStop}%
\bibitem [{\citenamefont {Vitanov}\ \emph {et~al.}(2001)\citenamefont
  {Vitanov}, \citenamefont {Halfmann}, \citenamefont {Shore},\ and\
  \citenamefont {Bergmann}}]{doi:10.1146/annurev.physchem.52.1.763}%
  \BibitemOpen
  \bibfield  {author} {\bibinfo {author} {\bibfnamefont {N.~V.}\ \bibnamefont
  {Vitanov}}, \bibinfo {author} {\bibfnamefont {T.}~\bibnamefont {Halfmann}},
  \bibinfo {author} {\bibfnamefont {B.~W.}\ \bibnamefont {Shore}}, \ and\
  \bibinfo {author} {\bibfnamefont {K.}~\bibnamefont {Bergmann}},\ }\bibfield
  {title} {\enquote {\bibinfo {title} {Laser-induced population transfer by
  adiabatic passage techniques},}\ }\href {\doibase
  10.1146/annurev.physchem.52.1.763} {\bibfield  {journal} {\bibinfo  {journal}
  {Annu. Rev. Phys. Chem.}\ }\textbf {\bibinfo {volume} {52}},\ \bibinfo
  {pages} {763--809} (\bibinfo {year} {2001})},\ \bibinfo {note} {pMID:
  11326080},\ \Eprint
  {http://arxiv.org/abs/http://dx.doi.org/10.1146/annurev.physchem.52.1.763}
  {http://dx.doi.org/10.1146/annurev.physchem.52.1.763} \BibitemShut {NoStop}%
\bibitem [{\citenamefont {Malinovsky}\ and\ \citenamefont
  {Krause}(2001)}]{PhysRevA.63.043415}%
  \BibitemOpen
  \bibfield  {author} {\bibinfo {author} {\bibfnamefont {V.~S.}\ \bibnamefont
  {Malinovsky}}\ and\ \bibinfo {author} {\bibfnamefont {J.~L.}\ \bibnamefont
  {Krause}},\ }\bibfield  {title} {\enquote {\bibinfo {title} {Efficiency and
  robustness of coherent population transfer with intense, chirped laser
  pulses},}\ }\href {\doibase 10.1103/PhysRevA.63.043415} {\bibfield  {journal}
  {\bibinfo  {journal} {Phys. Rev. A}\ }\textbf {\bibinfo {volume} {63}},\
  \bibinfo {pages} {043415} (\bibinfo {year} {2001})}\BibitemShut {NoStop}%
\bibitem [{\citenamefont {Chen}, \citenamefont {Tong},\ and\ \citenamefont
  {Mittra}(1997)}]{MOP:MOP14}%
  \BibitemOpen
  \bibfield  {author} {\bibinfo {author} {\bibfnamefont {Y.}~\bibnamefont
  {Chen}}, \bibinfo {author} {\bibfnamefont {M.-S.}\ \bibnamefont {Tong}}, \
  and\ \bibinfo {author} {\bibfnamefont {R.}~\bibnamefont {Mittra}},\
  }\bibfield  {title} {\enquote {\bibinfo {title} {Efficient and accurate
  finite-difference time-domain analysis of resonant structures using the
  blackman--harris window function},}\ }\href {\doibase
  10.1002/(SICI)1098-2760(19970820)15:6<389::AID-MOP14>3.0.CO;2-Y} {\bibfield
  {journal} {\bibinfo  {journal} {Microwave and Optical Technology Letters}\
  }\textbf {\bibinfo {volume} {15}},\ \bibinfo {pages} {389--392} (\bibinfo
  {year} {1997})}\BibitemShut {NoStop}%
\end{thebibliography}%

\end{document}